\RequirePackage{amsmath}
\documentclass[11pt,a4paper]{article}
\usepackage[margin=1in]{geometry}

\usepackage[x11names, svgnames, rgb]{xcolor}
\usepackage{tikz}
\usetikzlibrary{decorations,arrows,shapes}
\usetikzlibrary{calc,shapes,backgrounds}

\definecolor{strokecol}{rgb}{0.0,0.0,0.0}  %

\usepackage{soul}
\usepackage{graphicx}
\usepackage{enumitem}
\usepackage{amssymb}
\usepackage{amsthm}
\usepackage{multirow}
\usepackage{adjustbox}
\usepackage{subfigure}
\usepackage{xspace}

\newtheorem{example}{Example}

\usepackage[linesnumbered,ruled,vlined]{algorithm2e}
\DontPrintSemicolon%
\SetKwFunction{SAT}{SAT}
\SetKwFunction{tree}{tree}
\SetKwFunction{tenumerate}{TE}
\SetKwFunction{tdifference}{TDiff}
\SetKwFunction{tenumeratecard}{TE$^{\#}$}
\SetKwData{false}{false}\SetKwData{true}{true}
\SetKwInOut{Input}{input}
\SetKwInOut{Output}{output}
\SetKw{Func}{Function}

\usepackage[colorlinks=true,
            linkcolor=Blue4,
            citecolor=DarkGreen,
            urlcolor=Sienna4,
            filecolor=Sienna4,
            pdftitle={SAT-based Encodings for Optimal Decision Trees with Explicit Paths},
            pdfauthor={Mikol\'a\v{s} Janota, Ant\'onio Morgado}]{hyperref}
\def\miko{Mikol\'a\v{s} Janota}
\def\ajrm{Ant\'onio Morgado}

\title{SAT-based Encodings for Optimal Decision Trees with Explicit Paths}

\author{
    \miko\thanks{ORCID: 0000-0003-3487-784X}\\
    INESC-ID/IST, U.\ de Lisboa, Portugal\\
    Czech Technical University in Prague, Czech Republic
    \and
    \ajrm\thanks{ORCID: 0000-0002-5295-1321}\\
    INESC-ID/IST, U.\ de Lisboa, Portugal
}
\date{}

\newcommand\ninas{Narodytska~et~al.\xspace}
\newcommand\equal[2]{e^{#1}_{#2}}
\newcommand\go[2]{g^{#1}_{#2}}
\newcommand\terminate[2]{t^{#1}_{#2}}
\newcommand\assign[3]{a^{#1}_{#2,#3}}
\newcommand\match[2]{m^{#1}_{#2}}
\newcommand\class[1]{c^{#1}}
\newcommand\matchf[3]{m^{#1}_{#2,#3}}
\newcommand\maxpath{\textsf{P}}
\newcommand\maxstep{\textsf{S}}
\newcommand\maxfeature{\textsf{F}}
\newcommand\samples{\mathcal{E}}
\newcommand\lit{\text{lit}}
\newcommand\dtfinder{\texttt{dtfinder}\xspace}
\newcommand\dtfindertopo{\texttt{dtfinder-T}\xspace}
\newcommand\dtfindertopoord{\texttt{dtfinder-T-O}\xspace}
\newcommand\dtfindernarodytska{\texttt{dtfinder-DT1}\xspace}
\newcommand\dtfindernarodytskatopo{\texttt{dtfinder-DT1-T}\xspace}
\newcommand\dtfindernarodytskatopoord{\texttt{dtfinder-DT1-T-O}\xspace}
\newcommand\ddtfinder{\texttt{d-dtfinder}\xspace}
\newcommand\ddtfindertopord{\texttt{d-dtfinder-T-O}\xspace}
\newcommand\nina{\texttt{mindt}\xspace}
\newcommand\mindt{\texttt{mindt}\xspace}
\newcommand\sklearn{\texttt{sklearn}\xspace}

\newcommand\leaf{\ensuremath{\square}}
\newcommand\subtree{\ensuremath{\triangle}}

\begin{document}
\maketitle

\begin{abstract}
Decision trees play an important  role both in Machine Learning and Knowledge
Representation.  They are attractive  due to their immediate interpretability.
In the spirit of Occam's razor, and interpretability, it is desirable to
calculate the smallest tree.  This, however,  has proven to be a challenging
task and greedy approaches are typically used to learn trees in practice.
Nevertheless, recent work showed that by the use of SAT solvers one may
calculate the optimal size tree for real-world benchmarks.  This paper proposes
a novel SAT-based encoding that explicitly models paths in the tree, which
enables us to control the tree's depth as well as size.  At the level of
individual SAT calls, we investigate splitting the search space into tree
topologies. Our tool outperforms the existing implementation. But also,
the experimental results show that minimizing the depth first and then
minimizing the number of nodes enables solving a larger set of instances.
 \end{abstract}
\section{Introduction}\label{sec:Introduction}

Decision trees play an important role in machine learning either on their own~\cite{breiman-84}
or in the context of ensembles~\cite{breiman-ml01}.
Learning decision trees is especially attractive in the context of interpretable
machine learning due to their simplicity.  However, despite this simplicity,
 minimization of decision trees is well-known to be an NP-hard problem~\cite{hyafil-ipl76,hancock-ic96}.
  Yet, smaller trees are likely to   generalize better.

 To learn trees, suboptimal, greedy algorithms are used in practice.
With the rise of powerful reasoning engines, recent research  has
tackled the problem by the use of SAT, CSP, or MILP solvers~\cite{verwer-aaai19,verhaeghe-bnaic19,bertsimas-ml17,narodytska-ijcai18}.  Indeed, the
state-of-the-art technology shows that many (NP) hard problems are often
successfully solved. Conversely, such applications  drive the
reasoning technology by providing interesting benchmarks.

This paper, follows this line of research and proposes a novel SAT-based
encoding. This encoding enables finding a decision tree conforming to the
given set of examples with a given depth and number of nodes.
A minimal tree is found by iterative calls to a SAT solver while minimizing size and depth.

Focusing not only on size but also on depth of the tree brings about
opportunities for further analysis. Intuitively, more shallow trees are less likely to
over-fit. Indeed, modern packages such as Scikit~\cite{scikit} enable imposing a threshold
on the depth, which users have to set manually. Also, a shallow tree is
more likely to be  interpretable by a human because less memory is required to
keep track of a single branch.

The problem at hand is of challenging complexity. In practice, we may need to
deal with a high number of features and examples, which brings the search-space of possible trees into extreme dimensions. Looking for
an optimal tree means not only finding such tree but also proving that no
smaller tree exists.

The SAT technology has recently shown a lot of promise
in tackling difficult combinatorial  questions,
e.g.\ {E}rd{\H{o}}s' discrepancy~\cite{konev-ai15} or the Boolean Pythagorean
triples problem~\cite{heule-sat16}, among others.
Inspired by these results we also investigate the splitting of search-space based on the topology of the decision tree.
The paper has the following main contributions.

\begin{enumerate}
    \item It proposes a novel SAT-based encoding for decision trees,  along with a number of  optimizations.
    \item Compared to existing encoding, rather than representing nodes it represents \emph{paths} of the tree.
         This enables natively controlling not only the tree's size but also the tree's depth.
    \item It shows that minimizing depth first and then size enables tackling harder instances.
    \item It shows that search-space splitting by topologies enables tackling harder instances.
    \item The implemented  tool outperforms  existing work~\cite{narodytska-ijcai18}
\end{enumerate}
\section{Preliminaries}\label{sec:Preliminaries}
Standard notions and notation for propositional logic are assumed~\cite{DBLP:series/faia/SilvaLM09}.
A \emph{literal} is a Boolean variable ($x$)  or its negation (denoted $\lnot x$);
a \emph{clause} is a disjunction of literals a \emph{cube} is a conjunction of literals.
A formula is in \emph{conjunctive normal form (CNF)} if it is a conjunction of clauses.
General  Boolean formulas are also considered  constructed by using the standard connectives
conjunction ($\land$),
disjunction ($\lor$),
implication ($\rightarrow$),
bi-implication ($\leftrightarrow$).
State-of-the-art SAT solvers typically accept input in CNF.
Non-CNF  formulas  are converted to CNF by  standard equisatisfiable clausification methods~\cite{plaisted-jsc86}.

Several constraints in the paper also rely on \emph{cardinality constraints}~\cite{cardinality}.
These are also turned into CNF through standard means,  the implementation avails of the  cardinality encodings in the tool PySAT~\cite{mims-jsat15,ignatiev-sat18}.

\subsection {Training Data}
Standard setting of supervised learning is assumed~\cite{RussellNorvig10}.
Following notation and concepts of~\cite{narodytska-ijcai18}  we expect features to  be binary (with values 0, 1).
Non-binary features can be reduced to binary by unary or binary encoding.
Analogously,   classes are also binary (positive, negative).

Examples are defined on a fixed set of features $1..\maxfeature$
given as %
two sets, one containing the negative examples ($\samples^-$)
and second containing positive examples ($\samples^+$).
The examples are assumed consistent, i.e.\ $\samples^-\cap\samples^+ =\emptyset$.
 We write  $\samples$ for the whole set of examples, i.e.\ $\samples = \samples^-\cup\samples^+$.
Each example consists of feature-value pairs. We write $\sigma(q,f)$ for the value of a feature $f$ in an example $q$.
We assume that all the examples are complete, i.e.\ $\sigma(q)$ is total on $1..\maxfeature$.

\section{SAT-based Optimization of Decision Trees}\label{sec:Solving}
The objective is to develop a  propositional formula whose models  are
decision trees  congruent with the given set of samples.
Such model then  is found by a  call to an off-the-shelf  SAT solver.
As customary, we take the approach of optimizing by solving a series of decision problems.
This means finding a decision tree with a certain  size and diminishing  the size until no such tree exists.
Alternatively, other type of search can be used,  e.g., binary  or progression.

 This paper targets \emph{two}  optimization  criteria: \emph{size} and \emph{depth}.
Minimizing any combination of the two may be potentially be of interest.
Section~\ref{sec:Evaluation}  discusses the  exact type of search used in the implementation.

The structure of  binary trees guarantees a number of well-known properties.
Any tree with $n$ nodes has $(n+1)/2$  leaves and $(n-1)/2$ internal nodes.
Further, $n$ is always odd  and the number of leaves is equal to the number of
paths going from the root to a leaf.  Our encoding heavily exploits this
property:
\vspace{2pt}
\\
{\it Rather than modeling nodes of a tree, we model the set of unique paths from the root to leaves.}
\vspace{2pt}

The optimization algorithm has two levels. At the first level, search is being carried out on the tree's size and depth.
At the second level, the decision problem of finding a tree with such depth and size is solved via a SAT solver.
The SAT solver is used in a black-box fashion, i.e., the problem is encoded into its propositional form and  any off-the-shelf SAT solver may be used to solve it.

In the remainder of this section we focus on the decision problem, which is
invoked with
a given  number of paths~$\maxpath$ (controlling size) and maximum
allowed number of steps in a path~$\maxstep$ (controlling depth).

The steps  in a path are numbered in the following way. In the first step, each path is in the root. In the last step of a path,
the path goes from an internal node to a leaf. %
This means that if we  are looking for a tree with a particular depth and particular
 number of nodes we set $\maxstep$ and $\maxpath$ accordingly.  If we are
looking  only for a tree with minimal number of nodes but with an arbitrary depth,
the value of $\maxstep$ is set to  $\maxpath-1$,  which corresponds to the number of internal nodes.

\subsection{Path-based Encoding}\label{sec:Encoding}
The encoding we propose models each  path from the root to  a leaf separately while imposing relations between them that guarantee that the paths form a binary tree.
Throughout the paper, we use the convention that for a node labeled by a feature $f$,
the left child  corresponds to the value 0 of $f$
and the right child corresponds to the value 1 of $f$.

To model the tree, introduce a matrix
of  variables, where each row represents a path and each column represents a
step in the path.  The first row (the first path) is a path that only goes to
the left---it is the leftmost path in the tree.  Analogously, the last row (the
last path) is a path that only goes to the right---it is the rightmost path in
the tree.  In general, the paths are ordered in the way they would be obtained
by running DFS that goes to the left first.

Each path corresponds to  a sequence of 0's and 1's
so that 0 is a step to the left and 1 is a step to the
right.  Then, we consider these paths in a lexicographic order.  Each path is
represented by a sequence of variables, one for each step, where the variable
represents whether the path goes left or right in that step.  Additionally,
for each step we need to remember whether the  path has already  terminated and
which prefix is shared with the previous path.

\begin{table*}[t]
    \centering
  \begin{tabular}{|l|l|l|}
    \hline
      \textbf{Variable}  &  \textbf{Semantics}  & \textbf{Range} \\\hline
    $\go{p}{s}$  &  Path $p$ at step $s$ \textbf{g}oes right=1/left=0  &  $p\in 1..\maxpath,s\in 1..\maxstep$ \\\hline
    $\terminate{p}{s}$  &  Path $p$ at step $s$ is \textbf{t}erminated  &  $p\in 1..\maxpath,s\in 1..\maxstep+1$ \\\hline
    $\equal{p}{s}$  &  Path $p$ at step $s$ is \textbf{e}qual to path $p-1$  &  $p\in 2..\maxpath,s\in 1..\maxstep+1$ \\\hline
    \hline
    $\assign{p}{s}{f}$  &  Path $p$ at step $s$ is \textbf{a}ssigned feature $f$  &  $p\in 1..\maxpath,s\in 1..\maxstep, f\in 1..\maxfeature$\\\hline
    $\match{p}{q}$  &  Path $p$ \textbf{m}atches an example $q$  &  $p\in 1..\maxpath, q\in \samples$ \\\hline
       $\matchf{p}{f}{v}$  &  Path $p$ \textbf{m}atches on value $v$ for feature $f$  &  $p\in 1..\maxpath, f\in 1..\maxfeature, v\in \{0,1\}$\\\hline
       $\class{p}$  &  Path $p$ is \textbf{c}lassified as positive  &  $p\in 1..\maxpath$ \\\hline
   \end{tabular}
    \caption{Variables used in the encoding}\label {tab:variables}
\end{table*}

Table~\ref{tab:variables} summarizes the main variables  of the encoding.
The direction of each step $s$ in a path $p$ is determined by the variable~$\go{p}{s}$.
What is somewhat unusual about this encoding is that paths may share prefixes.
To that effect, the variable $e^p_s$
represents that the path $p$ in step $s$ is in the same node as the preceding path $p-1$.
The semantics of the variables $\equal{p}{s}$ is defined inductively. All paths share the root and therefore $\equal{p}{1}$ must be always true.
In further steps, paths $p$ and $p-1$ remain equal as long as both paths take steps in the same direction.

\begin{align}
    & \equal{p}{1}, p\in 2..\maxpath \label{eq:encoding:first}\\
    & \equal{p}{s+1} \leftrightarrow
    \left((\go{p}{s}\leftrightarrow\go{p-1}{s})\land\equal{p}{s}\right), p\in 2..\maxpath, s\in 1..\maxstep
\end{align}

Since it is unknown beforehand how many steps are in either path, the
variables $\terminate{p}{s}$ determine whether the path has
already terminated or not. Observe that the variables
$\terminate{p}{s}$  go up to step $\maxstep+1$,  whereas the variables
$\go{p}{s}$  go only  to step~$\maxstep$.  This is because  the $\go{p}{s}$
variables correspond to edges in the path while termination is tracked for nodes (as well as equality).
 A terminated path remains terminated  and cannot terminate if it is still equal to the previous one.
 Any path must be terminated after the last step.
\begin{align}
    & \terminate{p}{s}\rightarrow\terminate{p}{s+1}, p\in 1..\maxpath, s\in 1..\maxstep\\
    & \terminate{p}{s}\rightarrow\lnot\equal{p}{s}, p\in 2..\maxpath, s\in 1..\maxstep+1\\
    & \terminate{p}{\maxstep+1}, p\in 1..\maxpath
\end{align}

\begin{figure}[t]
    \centering
    \subfigure[example tree]{
        \centering
\tikzset{
  head/.style = {fill = white},
  trueleaf/.style = {fill = green!25, text = black, shape = rectangle},
  falseleaf/.style = {fill = red!25, text = black, shape = rectangle}
}

\begin{tikzpicture}[
    scale = 1, transform shape, thick,
    every node/.style = {draw, circle, minimum size = 3mm, inner sep=1pt},
    grow = down,  %
    level 1/.style = {sibling distance=2.5cm, level distance =0.7cm},
    level 2/.style = {sibling distance=1.5cm},
    level 3/.style = {sibling distance=4cm},
    level distance = .8cm
  ]
  \footnotesize
  \node[head] (Start)
          {$f_1$}
   child {   node [head] (A) {$f_2$}
     child { node [trueleaf] (B) {T}}
     child { node [falseleaf] (C) {F}}
   }
   child {   node [head] (D) {$f_3$}
     child { node [trueleaf] (E) {T}}
     child { node [falseleaf] (F) {F}}
   };

  \begin{scope}[nodes = {draw = none}]
    \path (Start) -- (A) node [near start, left]  {$0$};
    \path (A)     -- (B) node [near start, left]  {$0$};
    \path (A)     -- (C) node [near start, right] {$1$};
    \path (Start) -- (D) node [near start, right] {$1$};
    \path (D)     -- (E) node [near start, left]  {$0$};
    \path (D)     -- (F) node [near start, right] {$1$};
  \end{scope}
\end{tikzpicture}

     }
    \hspace{2pt}
    \subfigure[step direction]{
        \centering
        \begin{small}
            \begin{tabular}{|c|c|c|c|}
                \hline
                $g^p_s$ & \multicolumn{2}{c|}{$s$}\\
                \hline
                $p$ & 1 & 2 \\
                \hline
                1 & 0 & 0 \\
                2 & 0 & 1 \\
                3 & 1 & 0 \\
                4 & 1 & 1 \\
                \hline
            \end{tabular}
        \end{small}
    }
    \hspace{2pt}
    \subfigure[termination]{
        \centering
        \hspace{10pt}\begin{small}
            \begin{tabular}{|c|c|c|c|c|}
                \hline
                $t^p_s$ & \multicolumn{3}{c|}{$s$}\\
                \hline
                $p$ & 1 & 2 & 3\\
                \hline
                1 & 0 & 0 & 1 \\
                2 & 0 & 0 & 1 \\
                3 & 0 & 0 & 1 \\
                4 & 0 & 0 & 1 \\
                \hline
            \end{tabular}
        \end{small}\hspace{10pt}
    }
    \hspace{2pt}
    \subfigure[equality]{
        \centering\hspace{5pt}
        \begin{small}
            \begin{tabular}{|c|c|c|c|}
                \hline
                $e^p_s$ & \multicolumn{3}{c|}{$s$}\\
                \hline
                $p$  & $1$ & $2$ & $3$\\
                \hline
                $2$ & 1 & 1 & 0 \\
                $3$ & 1 & 0 & 0 \\
                $4$ & 1 & 1 & 0 \\
                \hline
            \end{tabular}
        \end{small}\hspace{5pt}
    }
    \caption{Assignment to the variables determining the tree's topology}\label{fig:ex:variables}
\end{figure}
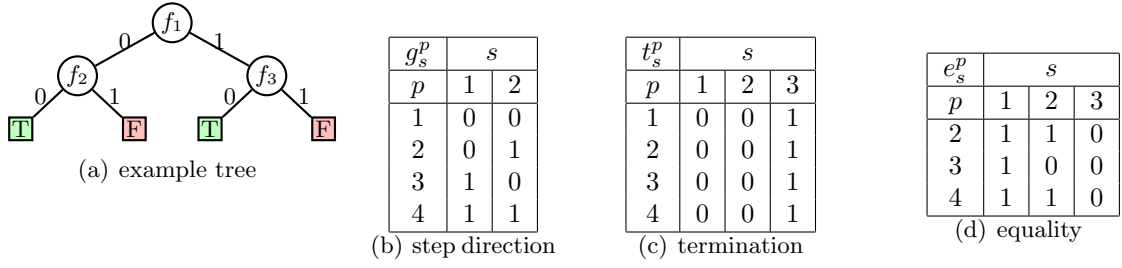

\begin{example}
    \autoref{fig:ex:variables} shows a binary tree  along with the values of
    the topology variables ($\go{p}{s}$, $\terminate{p}{s}$,  and
    $\equal{p}{s}$).  The tree is comprising 4 leaves, therefore 4 paths.  In
    this simple example each path makes two steps and then it terminates.  The
    second path shares everything with the first one except for the leaf.  The
    third path only shares the root with the second path.  The  last
    path shares everything with the third path, except for the leaf.  Observe
    that  since this is  a  full binary tree,  the $\go{p}{s}$ variables
    represent the binary numbers from 0 to 3.  %
\end{example}

Now  it is necessary to ensure that the  paths are lexicographically ordered.
The first path always goes left  and the last one always goes right.
If a  path $p$ in step $s$ is in the same node as path $p-1$, the path $p$ can go  left only
if $p-1$ also went left (otherwise they would cross).

\begin{align}
    & \lnot\go{1}{s}\land\go{\maxpath}{s}, s\in 1..\maxstep \\
    & \equal{p}{s}\rightarrow (\go{p-1}{s}\rightarrow\go{p}{s}), p\in 2..\maxpath, s\in 1..\maxstep+1
\end{align}

The lexicographic order alone does not guarantee  a correct topology.
Since the tree is binary, any path must adhere to the following pattern.
For a certain number of steps it shares the prefix with the preceding path until it  breaks off.
Once it breaks off, it has to go only to the left (or terminate).
At the same time, the preceding path can only go right  after  the break-off point (or terminate).
Otherwise, there would be a gap in the tree.
\begin{align}
    & (\lnot\terminate{p}{s}\land\lnot\equal{p}{s})\rightarrow \lnot\go{p}{s}, p\in 2..\maxpath, s\in 1..\maxstep\\
    & (\lnot\terminate{p}{s}\land\lnot\equal{p}{s})\rightarrow \go{p-1}{s}, p\in 2..\maxpath, s\in 1..\maxstep
\end{align}

\begin{figure}[t]
\begin{center}
\tikzset{
  head/.style = {fill = white},
  phantom/.style = {fill = none, draw = none},
  trueleaf/.style = {fill = green!25, text = black, shape = rectangle},
  falseleaf/.style = {fill = red!25, text = black, shape = rectangle}
}

\begin{tikzpicture}[
    yscale = 1, transform shape, thick,
    every node/.style = {draw, circle, solid, black, minimum size = 3mm, inner sep=1pt},
    grow = down,  %
    level 1/.style = {sibling distance=4cm, level distance = .75cm},
    level 2/.style = {sibling distance=3.5cm, level distance = .6cm},
    level 3/.style = {sibling distance=1.5cm},
    level 4/.style = {sibling distance=1cm},
    level 5/.style = {sibling distance=0.5cm}
  ]
  \footnotesize
  \node[head] (root) {$R$}
  child [dashed, green!80!black] { node[head] (A) {$A$}
    child [solid, black] { node [head] (B) {$B$}
     child [dotted] { node [phantom] (D) {}}
     child { node [head] (E) {$E$}
        child [dotted] { node [phantom] (H) {}}
        child { node [phantom] (I) {}}
     }
    }
    child [solid, black] { node [head] (C) {$C$}
     child { node [head] (F) {$F$}
      child { node [phantom] (J) {}}
      child [dotted] { node [phantom] (K) {}}
     }
     child [dotted] { node [phantom] (G) {}}
    }
    };

  \begin{scope}[nodes = {draw = none}]
    \draw [blue] (A) -- node [above, blue] {$0$} (B);
    \draw [red] (A) -- node [above, red] {$1$} (C) ;
    \draw [blue] (B) --node [above, near end,  blue] {$1$} (E) ;
    \draw [red] (C) --node [above, near end, red] {$0$} (F) ;
    \draw [blue] (E) -- (I) node [near start, right, blue] {$1$};
    \draw [red] (F) -- (J) node [near start, left, red] {$0$};

  \end{scope}
\end{tikzpicture}

 \caption{Two consecutive paths $R$--$E$ and $R$--$F$ diverging in node $A$.}
\label{fig:2pathstree}
\end{center}
\end{figure}
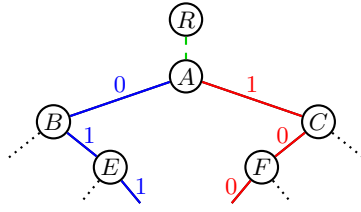

\autoref{fig:2pathstree} illustrates  these constraints.
Consider the blue path, $R\rightarrow A\rightarrow B\rightarrow E$
and the red path, $R\rightarrow A\rightarrow C\rightarrow F$,
where the blue one is lexicographically smaller.
The paths diverge in node $A$---blue goes left, the red goes right.
Afterwards, the blue path may only go right or terminate.
In contrast, the red path may only go left or terminate.  The
reason why this has to be the case is that for the red one to follow blue in
our ordering, the blue one has to  contain the \emph{last} path for the
subtree rooted in $B$ while the red one has to contain the  \emph{first}
path for the subtree rooted in $C$.

\subsubsection{Assigning Features and their Semantics}

The encoding of semantics of the training data is similar to the one in~\cite{narodytska-ijcai18}  but
 with two major differences:
 \begin{enumerate}
     \item  Here classification is only per \emph{path}, while
         in~\cite{narodytska-ijcai18} it is per \emph{node}  because any node
         can potentially be a leaf,
         which means semantics of the examples
         in our approach need only to be repeated  $(n+2)/2$  times rather than $n$ times.
  \item   Our encoding introduces explicit variables to track whether a given
      training example is matched for a given path, this is useful for one of
         the optimizations (see Section~\ref{sec:Optimizations}).
 \end{enumerate}

We make sure that each step is assigned exactly one feature and
that no feature appears more than once on any path.

\begin{align}
    \sum_{f\in 1..\maxfeature}\assign{p}{s}{f} = 1, p\in 1..\maxpath,s\in 1..\maxstep\\
    \sum_{p\in 1..\maxpath,s\in 1..\maxstep}\assign{p}{s}{f} \leq 1, f\in 1..\maxfeature
\end{align}

Recall that an example is seen as a set of feature-value pairs.
We say that a feature-value pair~$f,v$ is \emph{matched} on a path if the
path makes a step in the direction of~$v$ in the node  that is assigned the
feature~$f$.  An example is matched if all its feature-value pairs
are matched.  These two concepts are modeled by the variables
$\matchf{p}{f}{v}$ and $\match{p}{q} $, respectively.
Observe that $f,v$ is also matched  on any path that does not contain $f$ at all.
 Finally, once a path matches any positive example, it must be classified as positive and
the other way around.

\begin{align}
    & \matchf{p}{f}{0} \leftrightarrow \bigwedge_{s\in 1..\maxstep} (\lnot\terminate{p}{s}\land\assign{p}{s}{f}\rightarrow\lnot\go{p}{s}) & p\in 1..\maxpath, f\in 1..\maxfeature
    \\
    & \matchf{p}{f}{1} \leftrightarrow \bigwedge_{s\in 1..\maxstep} (\lnot\terminate{p}{s}\land\assign{p}{s}{f}\rightarrow\go{p}{s}) & p\in 1..\maxpath, f\in 1..\maxfeature
    \\
    &\match{p}{q} \leftrightarrow \bigwedge_{f\in 1..\maxfeature} \matchf{p}{f}{\sigma(q,f)} & p\in 1..\maxpath, q\in\samples
    \\
    &\match{p}{q} \rightarrow \class{p} & q\in\samples^+, p\in 1..\maxpath \\
    &\match{p}{q} \rightarrow\lnot\class{p} & q\in\samples^-, p\in 1..\maxpath\label{eq:encoding:last}
\end{align}

\paragraph{Summary of the encoding.}
The constraints \eqref{eq:encoding:first}--\eqref{eq:encoding:last} are
parameterized by natural numbers $\maxpath$ and $\maxstep$ and their satisfying
assignments represent a  sequence of $\maxpath$ paths in a binary tree from the root
to a leaf, where each path has at most $\maxstep$ edges.  The paths are
lexicographically ordered, starting from the leftmost path and ending in the
rightmost one. Additionally, the encoding ensures that there  are no  gaps between paths and
therefore  these represent the whole binary tree.  Each  node in a path
is labeled by a feature in a way that
 shared prefixes  among paths are labeled by the same features.
Each path is assigned a
classification class  that must be congruent with the training examples
given on the input.

\subsection{Path encoding Optimizations}\label{sec:Optimizations}
   The encoding described above permits  constructing any decision tree
   conforming to the given set of examples. However, certain optimizations can
   be made if we assume that we are not interested in superfluous nodes.

\subsubsection{Enforcing example matching}
   We make sure that any path (equivalently any leaf), \emph{matches} at least one
   of the given examples.  If it does not, its classification does not come from
   the examples and may therefore be arbitrary, which means it can be removed
   from any tree without violating the classification of the examples.
    At the formula level, additional constraints are added.
    \begin{align}
        \bigvee_{q\in\samples}\match{p}{q} &\quad p\in 1..\maxpath
    \end{align}

   \subsubsection{Pure features}
    Features with the same value in all the examples can be ignored as they
    never  permit distinguishing between two examples of  a different class. This is done at the preprocessing
    level, so the encoder never sees them.

   \subsubsection{Quasi-pure features}
   A feature may appear with a fixed value~$v$ within all the examples of one of
   the classes $c$. If such feature is assigned with the direction $v$,
   then the tree can immediately terminate with a leaf classified with $c$.
   As such, we enforce that the child of a node assigned in the direction of
   the value $v$ is a leaf classified with the class $c$.
   At the formula level, we add the following constraint
       for $s\in 1..\maxstep$ and $p\in 1..\maxpath$.
   \begin{align}
       (\assign{p}{s}{f}\land\lit(v, \go{p}{s}))\rightarrow (\terminate{p}{s+1}\land\lit(c, \class{p}))
       \text{ where } \lit(0, x) = \lnot x, \lit(1, x) = x
   \end{align}

   \subsubsection{Path lower bounds}
    We propose to use MaxSAT  to obtain lower bounds on the length of a path.
    The question we ask is what is the shortest  possible path that separates positive  and negatives examples.
    Since the lower bound considers only one path at a time, the order of features on that path is irrelevant.
     In preliminary experiments we have observed rather small lower  bounds.
    However,  the bound can be improved for the leftmost and rightmost branches.
    This gives us three types of bounds: for the leftmost and rightmost branches,
    and for any branch in between.
    In any path a feature either does not appear, or appears on a step that goes left or on a step that goes right.
    To  model this behavior we introduce  two variables  for each feature $x^0_f$ and $x^1_f$ (similar to the dual rail encoding~\cite{manquinho-ictai97}).
    This corresponds to the following hard  and soft constraints.

    \begin{align*}
        \text{hard:}\quad &\lnot x^0_f\lor\lnot x^1_f & f\in 1..\maxfeature\\
        \text{hard:}\quad &\lnot x^1_f / \lnot x^0_f & f\in 1..\maxfeature,\text{for leftmost/rightmost branch} \\
        \text{hard:}\quad &\bigvee_{f\in 1..\maxfeature} x^0_f \land \bigvee_{f\in 1..\maxfeature} x^1_f &\text{for general branch} \\
        \text{hard:}\quad &\match{p}{q} \leftrightarrow \bigwedge_{f\in 1..\maxfeature} \lnot x^{1-\sigma(q,f)} & p\in 1..\maxpath, q\in\samples\\
        \text{hard:}\quad &\bigwedge_{q\in\samples^+}\lnot\match{p}{q} \lor \bigwedge_{q\in\samples^-}\lnot\match{p}{q} &\\
        \text{soft:}\quad & \lnot x^v_f  & f\in 1..\maxfeature, v\in\{0,1\}
    \end{align*}

\section{Search-space Splitting by Topologies}\label{sec:topologies}

Upon  initial experiments,  we observed that the SAT  solver may
struggle even on decision trees of modest size, e.g.\ 9 nodes.  This is
somewhat surprising because the number of topologies does not initially grow  that much;
see Table~\ref{tab:Catalan}.

\begin{table}[t]
    \centering
    \begin{adjustbox}{width=\textwidth}
    \begin{tabular}{|c||c|c|c|c|c|c|c|c|c|c|c|c|c|c|c|}
        \hline
        $\mathbf{n}$ &    3   &  5   &  7   &  9   &  11   &  13   &  15   &  17   &  19   &  21   &  23   &  25   &  27   &  29   &  31\\\hline
        $\mathbf{t}$ &    1   &  2   &  5   &  14   &  42   &  132   &  429   &  1,430   &  4,862   &  16,796   &  58,786   &  208,012   &  742,900   &  2,674,440   &  9,694,845\\\hline
    \end{tabular}
    \end{adjustbox}
    \vspace{3pt}
    \caption{Number of topologies ($\mathbf{t}$) for tree size $\mathbf{n}\in 3..31$  (Catalan numbers)}\label{tab:Catalan}
\end{table}

This suggests splitting the search space into individual topologies and call
the SAT solver for each one of them separately. Like so, the SAT solver only
needs to find the labeling of the tree.  Intuitively this should be an easier problem because the SAT solver only needs to deal with one type of decisions.

This approach is not generally viable
because eventually the number of topologies is too large.  To which we propose
the following approach. The upper part of the topology is fixed---until a certain
depth---and the rest is left for the SAT solver to complete.  This gives rise
to  \emph{topology templates}. Each topology template is a  tree,
where each leaf is an actual leaf (\leaf) of the topology or an incomplete subtree (\subtree).

\begin{algorithm}[ht]
    \Func \tenumerate($n$, $d$)
    \Begin{%
        \lIf(\tcp*[f]{leaf}) {$n=1$}{\Return \{\leaf\}}
        \lIf(\tcp*[f]{incomplete subtree}) {$d=0$}{\Return \{\subtree\}}
        \If {$d = 1$}{%
            \lIf {$n=3$}{\Return \{ \tree(\leaf, \leaf) \}}
            \lElseIf {$n=5$}{\Return \{ \tree(\subtree, \leaf),  \tree(\leaf, \subtree) \}}
            \lElse {\Return \{ \tree(\leaf, \subtree), \tree(\subtree, \leaf), \tree(\subtree, \subtree) \}}
        }
        \Return $\{ \tree(l,r) \mid l \in \tenumerate(i, d-1), r \in \tenumerate(n-i-1, d-1), i \in 1..n-1 \}$\;
    }
    \caption{Topology enumeration, with
    \leaf\ -  leaf, \subtree\ - subtree}\label{algorithm:topology:enumeration}
\end{algorithm}

Algorithm~\ref{algorithm:topology:enumeration} recursively enumerates
incomplete  topologies on $n$ nodes with the cut-off parameter $d$.  In
order to avoid repetitions in enumeration, certain cases need to be treated
separately.  If the cut-off  parameter reaches 1, the children of the current
node will either be  leaves  (\leaf) or incomplete subtrees~(\subtree).  This,
in general gives three scenarios where either the left or the right child
is a leaf and the second child is a subtree, or both are subtrees. However, in
the case of~$n = 3$,  $n=5$  the scenarios are different.
Observe that because of  the cut-off parameter, the generated topology template may
have less than $n$ nodes.

We study topology enumeration both for our encoding as well as the encoding of \ninas~\cite{narodytska-ijcai18}.
A given topology template in the encoding of \ninas is enforced by a cube corresponding to the child relation and the
information whether a node is  a leaf or not.  An important property of our
generation procedure is that the cut-off  parameter is equal on all branches.  This
means that  numbering the topology template by BFS  gives the same numbers as
a BFS on any topology  corresponding to it. Since the  encoding of \nina
relies on BFS, this property lets us directly translate the relation into a cube.

\begin{algorithm}[b]
    \Func \tenumeratecard($n$, $d$)
    \Begin{%
        \lIf(\tcp*[f]{leaf}) {$n=1$}{\Return \{\leaf\}}
        \lIf(\tcp*[f]{incomplete  subtree of  size $n$}) {$d=0$}{\Return \{\#n\}}
        \Return $\{ \tree(l,r) \mid l \in \tenumeratecard(i, d-1), r \in \tenumeratecard(n-i-1, d-1), i \in 1..n-1 \}$\;
    }
    \caption{Topology enumeration with cardinalities}\label{algorithm:topologycard:enumeration}
\end{algorithm}

Our path-based encoding does not allow easily encoding a topology template because
the number of  paths in an incomplete  subtree (\subtree) is unknown.
 To this effect, we introduce a variation on the topology template where the leaves of the topology template are actual leaves (\leaf)
or an incomplete subtree with a given cardinality ($\#k$).
These topology templates can be easily enumerated as shown by  Algorithm~\ref{algorithm:topologycard:enumeration}.
Observe that the number of  these  topology templates  may be larger than in the previous version.
Such topology template is encoded into our path-based model in a straightforward fashion.
Each path in the topology template fixes
the direction in  prefixes in a  certain number of paths. The number of these paths corresponds to the  $\#k$ node at the end of the path.
Any path terminating in \leaf\ corresponds exactly to one path in the path-based model.

\subsection{Topology Enumeration}\label{sec:topologies:enumeration}

A cube describing a topology template  can either be encoded into assumptions to enable \emph{incremental
SAT solving}~\cite{een-entcs03} or appended as a set of unit clauses.
We observed that  in our case  incremental solving does not pay off for hard instances.  However,
at the same time, if a large number of  topology templates need to be examined,
initializing a new SAT solver for each one of them is too costly. Therefore,
the implementation employs both modes, incremental and non-incremental,
depending on the number of topology templates to be examined.

Another point of interest is the order in which the topology templates are
examined.  In the case of non-incremental SAT solving and unsatisfiable
instances, the order does not matter because all formulas need to be solved
independently of one another. Hence, the order plays mainly a role in the case
of satisfiable instances. The order heuristics we propose is the following.

We start with the assumption that we already have a suboptimal solution to the
problem  from a greedy  (fast) algorithm.  We would like to first focus on
topologies that are similar to the topology of this suboptimal solution.
In order to do so, we need some notion of \emph{difference} between topologies
(and topology templates).  For this purpose we define a simple function that
recursively compares the two topologies and accumulates a penalty once they are
different.  Additionally, subtrees with lower depth are accounted with less
weight.

\begin{algorithm}[t]
    \Func \tdifference$(t_1, t_2, w)$
    \Begin{%
        \lIf {$|t_1|=0$}{\Return $w|t_2|$}
        \lElseIf {$|t_2|=0$}{\Return $w|t_1|$}
        \lElse {\Return $ \tdifference(t_1.\text{left}, t_2.\text{left}, w\Delta) +
                            \tdifference(t_1.\text{right}, t_2.\text{right}, w\Delta) $}
    }
    \caption{Measuring difference between topologies}\label{algorithm:topology:difference}
\end{algorithm}

Algorithm~\ref{algorithm:topology:difference}
shows the function. If one of the given trees is empty, the penalty is the size of the other tree weighted by the factor~$w$.
Otherwise, the penalties are calculated as a sum of the left and the right subtrees,  respectively.
As the recursion descends, the weight is gradually decayed by the factor $\Delta\in (0,1]$.
In the implementation we chose the ad-hoc value of $0.75$.

 When partitioning the search space, the topology  templates are enumerated in the increasing order of the difference from the suboptimal solution.
\section{Experimental Evaluation}\label{sec:Evaluation}

The tool was  implemented  on top of the PySAT
package~\cite{ignatiev-sat18},  which interfaces with a number of modern SAT
solvers and provides a number of implementations of cardinality encodings.
We used the \emph{CaDiCaL solver}~\cite{cadical} and the \emph{$k$-Cardinality Modulo
Totalizer~\cite{mims-jsat15}}.  This configuration was chosen after some careful
preliminary experiments. We show that this configuration performs  significantly better
than the configuration used in the evaluation of Narodytska~et~al.

Our preliminary experiments also informed other ad-hoc choices that had to be made
as the search and encodings can be configured in a large number of ways.  An
alternative would be to employ automated parameter tuning in the spirit of
ParamILS~\cite{hutter2009paramils}; we leave this as future work.

The SAT solver is used in a non-incremental fashion, i.e.,  every decision
problem is solved independently of the other ones.  The exception is topology
enumeration: if the number of topologies is larger than  500,
the incremental mode is employed (see \autoref{sec:topologies}).

A suboptimal greedy solution is obtained by the popular modern machine learning
library \mbox{Scikit-learn~\cite{scikit}}, which also  enables a seamless integration
with the Python implementation.  The  greedy solution  is used in two
scenarios: 1)~to obtain an upper bound on the number of nodes in the solution
2)~to inform the ordering of topologies during  enumeration (see
Section~\ref{sec:topologies}).

The experiments were performed on servers with Intel(R) Xeon(R) CPU at
2.60GHz, 24 cores, 64GB RAM, while always running 4 tasks in parallel. The time
limit was set to 1000~seconds and the memory limit to~3~GB\@.
The experimental results report on the following search modes:
\begin{enumerate}[label={(\arabic*)}]
    \item binary search  on the number of nodes with no restriction on the depth  without  topology enumeration (with \sklearn upper-bound)\label{srch:abs}
    \item linearly increasing the number of nodes with no restriction on the depth with topology enumeration (linear UNSAT-SAT search)\label{srch:top}
    \item linearly increasing depth and linearly increasing number of nodes for each considered depth\label{srch:ddt}
\end{enumerate}

Searches~\ref{srch:abs} and~\ref{srch:top}
find the smallest tree just as in~\cite{narodytska-ijcai18}.
The search~\ref{srch:ddt} finds the smallest
tree in the lexicographic ordering  of the pair depth-size.

The evaluation was carried out on the benchmarks used
in~\cite{narodytska-ijcai18}, kindly provided by Narodytska.
These benchmarks were originally obtained by sampling a large set of instances~\cite{olsonCOUM17}, with sampling percentages of 20\%, and 50\% (we have used the same exact sampled benchmarks as Narodytska~et~al).
The reader is referred to~\cite{narodytska-ijcai18} for the details of the sampling procedure.

We compare our tools with the state-of-the-art tool \mindt~\cite{narodytska-ijcai18}.
Our tool is run in the following configurations.
The configuration~\dtfinder corresponds to the search~\ref{srch:abs}, i.e.\
size minimization via binary search and path-based encoding.
The configuration~\dtfindernarodytska is the same type of search but with encoding of Narodytska~et~al.
The suffix  \texttt{-T} in a configuration indicates topology-based search (search~\ref{srch:top}).
The suffix  \texttt{-T-O} topology-based is search with heuristic ordering.
The configuration~\ddtfinder corresponds to the search~\ref{srch:ddt}, i.e.\
depth-size minimization.

\begin{table*}[t]
\centering

\begin{tabular}{|c|c|c|c|c|c|c|c|c|}
\hline

 \% & \multicolumn{4}{c|}{\bf 0.2} & \multicolumn{4}{c|}{\bf 0.5} \\
 \hline
 \hline
 
 nf./ns. & \multicolumn{4}{c|}{447/136} & \multicolumn{4}{c|}{473/357} \\
 \hline
 \#I & \multicolumn{4}{c|}{754} & \multicolumn{4}{c|}{709} \\
 
\hline
\hline

& \#nd & depth  & cpu-time & \#slv & \#nd & depth & cpu-time & \#slv \\
\hline

\mindt & 6 & 3 & 58 & 394 & 5 & 3 & 52 & 249   \\
\hline
\hline

\dtfindernarodytska & 7 & 3 & 14 & 457 & 6 & 3 & 43 & 337 \\
\hline

\dtfindernarodytskatopo & 7 & 3 & 28 & 473 & 7 & 3 & 41 & 345 \\
\hline

\dtfindernarodytskatopoord & 7 & 3 & 27 & 473 & 7 & 3 & 39 & 345 \\
\hline
\hline

\dtfinder & 7 & 3 & 14 & 458 & 7 & 3 & 60 & 339 \\
\hline

\dtfindertopo & 7 & 3 & 29 & 470 & 7 & 3 & 42 & 342 \\
\hline

\dtfindertopoord & 7 & 3 & 30 & 471 & 7 & 3 & 39 & 341 \\
\hline
\hline

\ddtfinder & 8 & 3 & 65 & \textbf{519} & 7 & 3 & 73 & \textbf{352} \\
\hline

\ddtfindertopord & 8 & 3 & 49 & 486 & 7 & 3 & 57 & 345 \\
\hline
\hline

vbs & 8 & -- & 69 & 528 & 7 & -- & 46 & 355\\
\hline
 \end{tabular}
    \vspace{5pt}
\caption{Results on all the benchmarks divided by the percentage of random sampling.}
\label{tab:allresults}
\end{table*}
 
\begin{figure}
    \centering
    \includegraphics[width=0.75\textwidth]{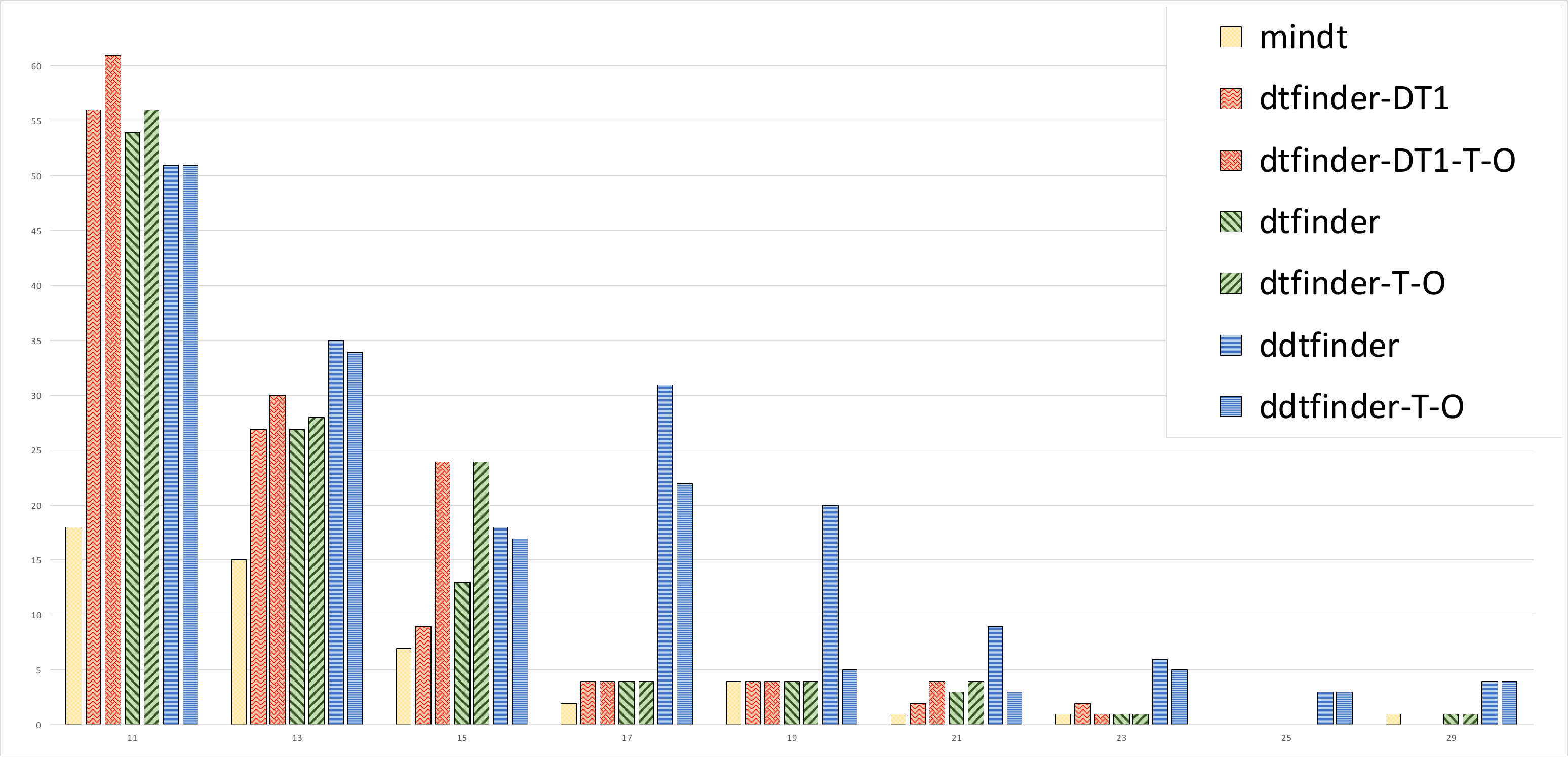}
    \caption{Distribution of the sizes of  calculated optimal trees}\label{fig:histogram}
\end{figure}

\autoref{tab:allresults} summarizes results for all the considered benchmarks  and tools.
The first row (\%) shows the percentage of random samplings  used to construct the  instance,
the second row the average number of features (nf.) and samples (ns.).
The third row (\#I) shows the number of benchmarks in that category.
The remaining rows are grouped according to the tool they represent.

For each of the tools we present four values: the average number of nodes discovered (\#nd); the average depth of the tree reported (depth); the average CPU time taken in solved instances (cpu-time); and the number of instances solved (\#slv).

\autoref{fig:histogram} shows a histogram of the sizes of the optimal trees per
solver. The vertical axis shows the number of solved instances and the  horizontal groups
the solvers according to the number of nodes of the reported decision trees.
More detailed overview of the data can be found here on the authors' website~\cite{eval}.

\autoref{tab:allresults} enables the following conclusions.
Our implementation (\dtfinder) outperforms  the tool by \ninas (\mindt) in all cases.
This is also the case  for  their encoding; We attribute this to the choice of cardinality
 encoding and the  SAT solver. We used
$k$-Cardinality Modulo Totalizer and CaDiCaL
while \mindt  uses  sequential counter and glucose-0.3.

Comparing \dtfinder with \ddtfinder, we can see that \ddtfinder is faster to
compute a minimum depth solution than \dtfinder is to compute a minimum size
solution and even more interestingly, again solves even more instances.

The  topology search-space splitting is beneficial in all encodings except for depth minimization.
Both our encoding and encoding of \ninas  solves more instances with topology enumeration.
Not always this helps the average CPU time;  however, it went from 60s to 39s in
path-based  encoding for the 0.5 instances.
The ordering of topologies enables a minor speed-up but overall the effect is small.

The distribution of sizes of solved instances (\autoref{fig:histogram})
shows that the hardness of an instance grows drastically with the size.
While depth minimization is able to solve a handful of instances of size 29,
the path-based encoding solves just 1, our implementation of \ninas none and, surprisingly
\mindt 1. This can be attributed to the number of topologies (see \autoref{tab:Catalan}).

\begin{figure}[t]
    \centering
    \subfigure[\ddtfinder]{%

\begin{tikzpicture}[>=latex',line join=bevel,xscale=.2,yscale=0.13,font=\tiny]
\pgfsetcolor{black}
  \pgfsetcolor{blue}
  \draw [->] (162.51bp,435.39bp) .. controls (157.35bp,423.25bp) and (150.38bp,406.87bp)  .. (140.48bp,383.58bp);
  \pgfsetstrokecolor{strokecol}
  \draw (157.5bp,409.5bp) node {0};
  \pgfsetcolor{red}
  \draw [->] (177.49bp,435.39bp) .. controls (182.65bp,423.25bp) and (189.62bp,406.87bp)  .. (199.52bp,383.58bp);
  \pgfsetstrokecolor{strokecol}
  \draw (194.5bp,409.5bp) node {1};
  \pgfsetcolor{blue}
  \draw [->] (119.65bp,349.64bp) .. controls (109.44bp,337.13bp) and (95.161bp,319.63bp)  .. (76.765bp,297.09bp);
  \pgfsetstrokecolor{strokecol}
  \draw (105.5bp,322.5bp) node {0};
  \pgfsetcolor{red}
  \draw [->] (133.21bp,347.97bp) .. controls (133.34bp,336.19bp) and (133.52bp,320.56bp)  .. (133.79bp,297.0bp);
  \pgfsetstrokecolor{strokecol}
  \draw (137.5bp,322.5bp) node {1};
  \pgfsetcolor{blue}
  \draw [->] (207.0bp,347.97bp) .. controls (207.0bp,336.19bp) and (207.0bp,320.56bp)  .. (207.0bp,297.0bp);
  \pgfsetstrokecolor{strokecol}
  \draw (210.5bp,322.5bp) node {0};
  \pgfsetcolor{red}
  \draw [->] (229.06bp,355.16bp) .. controls (260.23bp,339.84bp) and (317.32bp,311.77bp)  .. (362.22bp,289.71bp);
  \pgfsetstrokecolor{strokecol}
  \draw (310.5bp,322.5bp) node {1};
  \pgfsetcolor{blue}
  \draw [->] (189.65bp,265.02bp) .. controls (171.93bp,250.75bp) and (144.26bp,228.46bp)  .. (116.31bp,205.94bp);
  \pgfsetstrokecolor{strokecol}
  \draw (163.5bp,235.5bp) node {0};
  \pgfsetcolor{red}
  \draw [->] (207.0bp,260.97bp) .. controls (207.0bp,249.19bp) and (207.0bp,233.56bp)  .. (207.0bp,210.0bp);
  \pgfsetstrokecolor{strokecol}
  \draw (210.5bp,235.5bp) node {1};
  \pgfsetcolor{blue}
  \draw [->] (85.794bp,176.04bp) .. controls (75.381bp,163.46bp) and (60.645bp,145.65bp)  .. (42.115bp,123.26bp);
  \pgfsetstrokecolor{strokecol}
  \draw (71.5bp,148.5bp) node {0};
  \pgfsetcolor{red}
  \draw [->] (99.0bp,173.97bp) .. controls (99.0bp,162.19bp) and (99.0bp,146.56bp)  .. (99.0bp,123.0bp);
  \pgfsetstrokecolor{strokecol}
  \draw (102.5bp,148.5bp) node {1};
  \pgfsetcolor{blue}
  \draw [->] (200.12bp,174.39bp) .. controls (195.37bp,162.25bp) and (188.97bp,145.87bp)  .. (179.87bp,122.58bp);
  \pgfsetstrokecolor{strokecol}
  \draw (196.5bp,148.5bp) node {0};
  \pgfsetcolor{red}
  \draw [->] (216.29bp,174.81bp) .. controls (222.94bp,162.5bp) and (232.02bp,145.68bp)  .. (244.57bp,122.45bp);
  \pgfsetstrokecolor{strokecol}
  \draw (237.5bp,148.5bp) node {1};
  \pgfsetcolor{blue}
  \draw [->] (159.46bp,88.636bp) .. controls (149.11bp,76.127bp) and (134.63bp,58.634bp)  .. (115.97bp,36.092bp);
  \pgfsetstrokecolor{strokecol}
  \draw (145.5bp,61.5bp) node {0};
  \pgfsetcolor{red}
  \draw [->] (173.0bp,86.974bp) .. controls (173.0bp,75.192bp) and (173.0bp,59.561bp)  .. (173.0bp,36.003bp);
  \pgfsetstrokecolor{strokecol}
  \draw (176.5bp,61.5bp) node {1};
  \pgfsetcolor{blue}
  \draw [->] (252.14bp,86.974bp) .. controls (250.92bp,75.192bp) and (249.3bp,59.561bp)  .. (246.86bp,36.003bp);
  \pgfsetstrokecolor{strokecol}
  \draw (254.5bp,61.5bp) node {0};
  \pgfsetcolor{red}
  \draw [->] (266.45bp,87.812bp) .. controls (275.42bp,75.427bp) and (287.68bp,58.484bp)  .. (303.93bp,36.051bp);
  \pgfsetstrokecolor{strokecol}
  \draw (293.5bp,61.5bp) node {1};
  \pgfsetcolor{blue}
  \draw [->] (384.0bp,260.97bp) .. controls (384.0bp,249.19bp) and (384.0bp,233.56bp)  .. (384.0bp,210.0bp);
  \pgfsetstrokecolor{strokecol}
  \draw (387.5bp,235.5bp) node {0};
  \pgfsetcolor{red}
  \draw [->] (406.07bp,268.51bp) .. controls (438.01bp,253.32bp) and (497.33bp,225.12bp)  .. (543.69bp,203.08bp);
  \pgfsetstrokecolor{strokecol}
  \draw (490.5bp,235.5bp) node {1};
  \pgfsetcolor{red}
  \draw [->] (378.41bp,173.97bp) .. controls (374.71bp,162.08bp) and (369.8bp,146.25bp)  .. (362.59bp,123.0bp);
  \pgfsetstrokecolor{strokecol}
  \draw (376.5bp,148.5bp) node {1};
  \pgfsetcolor{blue}
  \draw [->] (394.27bp,174.81bp) .. controls (401.75bp,162.3bp) and (412.01bp,145.14bp)  .. (425.75bp,122.15bp);
  \pgfsetstrokecolor{strokecol}
  \draw (417.5bp,148.5bp) node {0};
  \pgfsetcolor{blue}
  \draw [->] (426.49bp,87.394bp) .. controls (419.95bp,75.292bp) and (411.14bp,58.985bp)  .. (398.82bp,36.178bp);
  \pgfsetstrokecolor{strokecol}
  \draw (419.5bp,61.5bp) node {0};
  \pgfsetcolor{red}
  \draw [->] (441.18bp,86.974bp) .. controls (444.6bp,75.075bp) and (449.15bp,59.251bp)  .. (455.83bp,36.003bp);
  \pgfsetstrokecolor{strokecol}
  \draw (453.5bp,61.5bp) node {1};
  \pgfsetcolor{blue}
  \draw [->] (567.0bp,173.97bp) .. controls (567.0bp,162.19bp) and (567.0bp,146.56bp)  .. (567.0bp,123.0bp);
  \pgfsetstrokecolor{strokecol}
  \draw (570.5bp,148.5bp) node {0};
  \pgfsetcolor{red}
  \draw [->] (585.16bp,177.63bp) .. controls (603.08bp,163.47bp) and (630.64bp,141.67bp)  .. (658.81bp,119.39bp);
  \pgfsetstrokecolor{strokecol}
  \draw (632.5bp,148.5bp) node {1};
  \pgfsetcolor{blue}
  \draw [->] (560.12bp,87.394bp) .. controls (555.44bp,75.408bp) and (549.14bp,59.298bp)  .. (540.1bp,36.178bp);
  \pgfsetstrokecolor{strokecol}
  \draw (556.5bp,61.5bp) node {0};
  \pgfsetcolor{red}
  \draw [->] (574.69bp,87.394bp) .. controls (579.93bp,75.408bp) and (586.96bp,59.298bp)  .. (597.06bp,36.178bp);
  \pgfsetstrokecolor{strokecol}
  \draw (592.5bp,61.5bp) node {1};
  \pgfsetcolor{blue}
  \draw [->] (677.0bp,86.974bp) .. controls (677.0bp,75.192bp) and (677.0bp,59.561bp)  .. (677.0bp,36.003bp);
  \pgfsetstrokecolor{strokecol}
  \draw (680.5bp,61.5bp) node {0};
  \pgfsetcolor{red}
  \draw [->] (690.54bp,88.636bp) .. controls (700.89bp,76.127bp) and (715.37bp,58.634bp)  .. (734.03bp,36.092bp);
  \pgfsetstrokecolor{strokecol}
  \draw (721.5bp,61.5bp) node {1};
\begin{scope}
  \pgfsetstrokecolor{strokecol}
  \draw (89.0bp,297.0bp) -- (35.0bp,297.0bp) -- (35.0bp,261.0bp) -- (89.0bp,261.0bp) -- cycle;
  \draw (62.0bp,279.0bp) node {F};
\end{scope}
\begin{scope}
  \pgfsetstrokecolor{strokecol}
  \draw (161.0bp,297.0bp) -- (107.0bp,297.0bp) -- (107.0bp,261.0bp) -- (161.0bp,261.0bp) -- cycle;
  \draw (134.0bp,279.0bp) node {T};
\end{scope}
\begin{scope}
  \pgfsetstrokecolor{strokecol}
  \draw (54.0bp,123.0bp) -- (0.0bp,123.0bp) -- (0.0bp,87.0bp) -- (54.0bp,87.0bp) -- cycle;
  \draw (27.0bp,105.0bp) node {T};
\end{scope}
\begin{scope}
  \pgfsetstrokecolor{strokecol}
  \draw (126.0bp,123.0bp) -- (72.0bp,123.0bp) -- (72.0bp,87.0bp) -- (126.0bp,87.0bp) -- cycle;
  \draw (99.0bp,105.0bp) node {F};
\end{scope}
\begin{scope}
  \pgfsetstrokecolor{strokecol}
  \draw (128.0bp,36.0bp) -- (74.0bp,36.0bp) -- (74.0bp,0.0bp) -- (128.0bp,0.0bp) -- cycle;
  \draw (101.0bp,18.0bp) node {T};
\end{scope}
\begin{scope}
  \pgfsetstrokecolor{strokecol}
  \draw (200.0bp,36.0bp) -- (146.0bp,36.0bp) -- (146.0bp,0.0bp) -- (200.0bp,0.0bp) -- cycle;
  \draw (173.0bp,18.0bp) node {F};
\end{scope}
\begin{scope}
  \pgfsetstrokecolor{strokecol}
  \draw (272.0bp,36.0bp) -- (218.0bp,36.0bp) -- (218.0bp,0.0bp) -- (272.0bp,0.0bp) -- cycle;
  \draw (245.0bp,18.0bp) node {F};
\end{scope}
\begin{scope}
  \pgfsetstrokecolor{strokecol}
  \draw (344.0bp,36.0bp) -- (290.0bp,36.0bp) -- (290.0bp,0.0bp) -- (344.0bp,0.0bp) -- cycle;
  \draw (317.0bp,18.0bp) node {T};
\end{scope}
\begin{scope}
  \pgfsetstrokecolor{strokecol}
  \draw (416.0bp,36.0bp) -- (362.0bp,36.0bp) -- (362.0bp,0.0bp) -- (416.0bp,0.0bp) -- cycle;
  \draw (389.0bp,18.0bp) node {T};
\end{scope}
\begin{scope}
  \pgfsetstrokecolor{strokecol}
  \draw (488.0bp,36.0bp) -- (434.0bp,36.0bp) -- (434.0bp,0.0bp) -- (488.0bp,0.0bp) -- cycle;
  \draw (461.0bp,18.0bp) node {F};
\end{scope}
\begin{scope}
  \pgfsetstrokecolor{strokecol}
  \draw (384.0bp,123.0bp) -- (330.0bp,123.0bp) -- (330.0bp,87.0bp) -- (384.0bp,87.0bp) -- cycle;
  \draw (357.0bp,105.0bp) node {T};
\end{scope}
\begin{scope}
  \pgfsetstrokecolor{strokecol}
  \draw (560.0bp,36.0bp) -- (506.0bp,36.0bp) -- (506.0bp,0.0bp) -- (560.0bp,0.0bp) -- cycle;
  \draw (533.0bp,18.0bp) node {F};
\end{scope}
\begin{scope}
  \pgfsetstrokecolor{strokecol}
  \draw (632.0bp,36.0bp) -- (578.0bp,36.0bp) -- (578.0bp,0.0bp) -- (632.0bp,0.0bp) -- cycle;
  \draw (605.0bp,18.0bp) node {T};
\end{scope}
\begin{scope}
  \pgfsetstrokecolor{strokecol}
  \draw (704.0bp,36.0bp) -- (650.0bp,36.0bp) -- (650.0bp,0.0bp) -- (704.0bp,0.0bp) -- cycle;
  \draw (677.0bp,18.0bp) node {T};
\end{scope}
\begin{scope}
  \pgfsetstrokecolor{strokecol}
  \draw (776.0bp,36.0bp) -- (722.0bp,36.0bp) -- (722.0bp,0.0bp) -- (776.0bp,0.0bp) -- cycle;
  \draw (749.0bp,18.0bp) node {F};
\end{scope}
\begin{scope}
  \pgfsetstrokecolor{strokecol}
  \draw (170.0bp,453.0bp) ellipse (29.5bp and 18.0bp);
  \draw (170.0bp,453.0bp) node {1};
\end{scope}
\begin{scope}
  \pgfsetstrokecolor{strokecol}
  \draw (133.0bp,366.0bp) ellipse (29.5bp and 18.0bp);
  \draw (133.0bp,366.0bp) node {2};
\end{scope}
\begin{scope}
  \pgfsetstrokecolor{strokecol}
  \draw (207.0bp,366.0bp) ellipse (27.0bp and 18.0bp);
  \draw (207.0bp,366.0bp) node {3};
\end{scope}
\begin{scope}
  \pgfsetstrokecolor{strokecol}
  \draw (207.0bp,279.0bp) ellipse (27.0bp and 18.0bp);
  \draw (207.0bp,279.0bp) node {6};
\end{scope}
\begin{scope}
  \pgfsetstrokecolor{strokecol}
  \draw (384.0bp,279.0bp) ellipse (27.0bp and 18.0bp);
  \draw (384.0bp,279.0bp) node {7};
\end{scope}
\begin{scope}
  \pgfsetstrokecolor{strokecol}
  \draw (99.0bp,192.0bp) ellipse (27.0bp and 18.0bp);
  \draw (99.0bp,192.0bp) node {8};
\end{scope}
\begin{scope}
  \pgfsetstrokecolor{strokecol}
  \draw (207.0bp,192.0bp) ellipse (29.5bp and 18.0bp);
  \draw (207.0bp,192.0bp) node {9};
\end{scope}
\begin{scope}
  \pgfsetstrokecolor{strokecol}
  \draw (173.0bp,105.0bp) ellipse (29.5bp and 18.0bp);
  \draw (173.0bp,105.0bp) node {12};
\end{scope}
\begin{scope}
  \pgfsetstrokecolor{strokecol}
  \draw (254.0bp,105.0bp) ellipse (33.6bp and 18.0bp);
  \draw (254.0bp,105.0bp) node {13};
\end{scope}
\begin{scope}
  \pgfsetstrokecolor{strokecol}
  \draw (384.0bp,192.0bp) ellipse (33.6bp and 18.0bp);
  \draw (384.0bp,192.0bp) node {18};
\end{scope}
\begin{scope}
  \pgfsetstrokecolor{strokecol}
  \draw (567.0bp,192.0bp) ellipse (29.5bp and 18.0bp);
  \draw (567.0bp,192.0bp) node {19};
\end{scope}
\begin{scope}
  \pgfsetstrokecolor{strokecol}
  \draw (436.0bp,105.0bp) ellipse (33.6bp and 18.0bp);
  \draw (436.0bp,105.0bp) node {20};
\end{scope}
\begin{scope}
  \pgfsetstrokecolor{strokecol}
  \draw (567.0bp,105.0bp) ellipse (29.5bp and 18.0bp);
  \draw (567.0bp,105.0bp) node {24};
\end{scope}
\begin{scope}
  \pgfsetstrokecolor{strokecol}
  \draw (677.0bp,105.0bp) ellipse (29.5bp and 18.0bp);
  \draw (677.0bp,105.0bp) node {25};
\end{scope}
\end{tikzpicture}

}
    \subfigure[\sklearn]{%

\begin{tikzpicture}[>=latex',line join=bevel,xscale=.2,yscale=0.13,font=\tiny]
\pgfsetcolor{black}
  \pgfsetcolor{blue}
  \draw [->] (355.91bp,870.39bp) .. controls (350.27bp,858.13bp) and (342.64bp,841.55bp)  .. (332.08bp,818.58bp);
  \pgfsetstrokecolor{strokecol}
  \draw (350.5bp,844.5bp) node {0};
  \pgfsetcolor{red}
  \draw [->] (372.1bp,870.81bp) .. controls (377.86bp,858.58bp) and (385.72bp,841.9bp)  .. (396.64bp,818.75bp);
  \pgfsetstrokecolor{strokecol}
  \draw (390.5bp,844.5bp) node {1};
  \pgfsetcolor{blue}
  \draw [->] (306.19bp,786.24bp) .. controls (289.09bp,772.07bp) and (263.1bp,750.54bp)  .. (236.2bp,728.25bp);
  \pgfsetstrokecolor{strokecol}
  \draw (281.5bp,757.5bp) node {0};
  \pgfsetcolor{red}
  \draw [->] (324.41bp,782.97bp) .. controls (324.69bp,771.19bp) and (325.04bp,755.56bp)  .. (325.59bp,732.0bp);
  \pgfsetstrokecolor{strokecol}
  \draw (329.5bp,757.5bp) node {1};
  \pgfsetcolor{red}
  \draw [->] (205.43bp,698.04bp) .. controls (194.72bp,685.46bp) and (179.58bp,667.65bp)  .. (160.54bp,645.26bp);
  \pgfsetstrokecolor{strokecol}
  \draw (190.5bp,670.5bp) node {1};
  \pgfsetcolor{blue}
  \draw [->] (219.0bp,695.97bp) .. controls (219.0bp,684.19bp) and (219.0bp,668.56bp)  .. (219.0bp,645.0bp);
  \pgfsetstrokecolor{strokecol}
  \draw (222.5bp,670.5bp) node {0};
  \pgfsetcolor{red}
  \draw [->] (205.46bp,610.64bp) .. controls (195.11bp,598.13bp) and (180.63bp,580.63bp)  .. (161.97bp,558.09bp);
  \pgfsetstrokecolor{strokecol}
  \draw (191.5bp,583.5bp) node {1};
  \pgfsetcolor{blue}
  \draw [->] (219.0bp,608.97bp) .. controls (219.0bp,597.19bp) and (219.0bp,581.56bp)  .. (219.0bp,558.0bp);
  \pgfsetstrokecolor{strokecol}
  \draw (222.5bp,583.5bp) node {0};
  \pgfsetcolor{blue}
  \draw [->] (209.91bp,522.81bp) .. controls (203.36bp,510.42bp) and (194.39bp,493.46bp)  .. (182.07bp,470.15bp);
  \pgfsetstrokecolor{strokecol}
  \draw (202.5bp,496.5bp) node {0};
  \pgfsetcolor{red}
  \draw [->] (225.07bp,522.39bp) .. controls (229.23bp,510.33bp) and (234.83bp,494.08bp)  .. (242.84bp,470.88bp);
  \pgfsetstrokecolor{strokecol}
  \draw (239.5bp,496.5bp) node {1};
  \pgfsetcolor{blue}
  \draw [->] (159.46bp,436.64bp) .. controls (148.53bp,423.43bp) and (133.0bp,404.67bp)  .. (114.24bp,382.0bp);
  \pgfsetstrokecolor{strokecol}
  \draw (145.5bp,409.5bp) node {0};
  \pgfsetcolor{red}
  \draw [->] (173.41bp,434.97bp) .. controls (173.69bp,423.19bp) and (174.04bp,407.56bp)  .. (174.59bp,384.0bp);
  \pgfsetstrokecolor{strokecol}
  \draw (178.5bp,409.5bp) node {1};
  \pgfsetcolor{red}
  \draw [->] (87.427bp,350.04bp) .. controls (76.725bp,337.46bp) and (61.58bp,319.65bp)  .. (42.535bp,297.26bp);
  \pgfsetstrokecolor{strokecol}
  \draw (72.5bp,322.5bp) node {1};
  \pgfsetcolor{blue}
  \draw [->] (101.0bp,347.97bp) .. controls (101.0bp,336.19bp) and (101.0bp,320.56bp)  .. (101.0bp,297.0bp);
  \pgfsetstrokecolor{strokecol}
  \draw (104.5bp,322.5bp) node {0};
  \pgfsetcolor{blue}
  \draw [->] (95.613bp,260.97bp) .. controls (92.057bp,249.08bp) and (87.328bp,233.25bp)  .. (80.38bp,210.0bp);
  \pgfsetstrokecolor{strokecol}
  \draw (93.5bp,235.5bp) node {0};
  \pgfsetcolor{red}
  \draw [->] (110.48bp,261.81bp) .. controls (117.39bp,249.3bp) and (126.86bp,232.14bp)  .. (139.54bp,209.15bp);
  \pgfsetstrokecolor{strokecol}
  \draw (131.5bp,235.5bp) node {1};
  \pgfsetcolor{red}
  \draw [->] (143.41bp,173.97bp) .. controls (139.71bp,162.08bp) and (134.8bp,146.25bp)  .. (127.59bp,123.0bp);
  \pgfsetstrokecolor{strokecol}
  \draw (141.5bp,148.5bp) node {1};
  \pgfsetcolor{blue}
  \draw [->] (159.27bp,174.81bp) .. controls (166.75bp,162.3bp) and (177.01bp,145.14bp)  .. (190.75bp,122.15bp);
  \pgfsetstrokecolor{strokecol}
  \draw (182.5bp,148.5bp) node {0};
  \pgfsetcolor{blue}
  \draw [->] (193.71bp,87.394bp) .. controls (188.76bp,75.408bp) and (182.09bp,59.298bp)  .. (172.52bp,36.178bp);
  \pgfsetstrokecolor{strokecol}
  \draw (189.5bp,61.5bp) node {0};
  \pgfsetcolor{red}
  \draw [->] (208.29bp,87.394bp) .. controls (213.24bp,75.408bp) and (219.91bp,59.298bp)  .. (229.48bp,36.178bp);
  \pgfsetstrokecolor{strokecol}
  \draw (225.5bp,61.5bp) node {1};
  \pgfsetcolor{blue}
  \draw [->] (175.0bp,347.97bp) .. controls (175.0bp,336.19bp) and (175.0bp,320.56bp)  .. (175.0bp,297.0bp);
  \pgfsetstrokecolor{strokecol}
  \draw (178.5bp,322.5bp) node {0};
  \pgfsetcolor{red}
  \draw [->] (188.54bp,349.64bp) .. controls (198.89bp,337.13bp) and (213.37bp,319.63bp)  .. (232.03bp,297.09bp);
  \pgfsetstrokecolor{strokecol}
  \draw (219.5bp,322.5bp) node {1};
  \pgfsetcolor{blue}
  \draw [->] (249.0bp,434.97bp) .. controls (249.0bp,423.19bp) and (249.0bp,407.56bp)  .. (249.0bp,384.0bp);
  \pgfsetstrokecolor{strokecol}
  \draw (252.5bp,409.5bp) node {0};
  \pgfsetcolor{red}
  \draw [->] (262.57bp,437.04bp) .. controls (273.77bp,423.88bp) and (289.82bp,405.0bp)  .. (309.26bp,382.15bp);
  \pgfsetstrokecolor{strokecol}
  \draw (294.5bp,409.5bp) node {1};
  \pgfsetcolor{blue}
  \draw [->] (322.59bp,347.97bp) .. controls (322.31bp,336.19bp) and (321.96bp,320.56bp)  .. (321.41bp,297.0bp);
  \pgfsetstrokecolor{strokecol}
  \draw (326.5bp,322.5bp) node {0};
  \pgfsetcolor{red}
  \draw [->] (336.17bp,349.64bp) .. controls (346.23bp,337.13bp) and (360.31bp,319.63bp)  .. (378.44bp,297.09bp);
  \pgfsetstrokecolor{strokecol}
  \draw (366.5bp,322.5bp) node {1};
  \pgfsetcolor{blue}
  \draw [->] (326.0bp,695.97bp) .. controls (326.0bp,684.19bp) and (326.0bp,668.56bp)  .. (326.0bp,645.0bp);
  \pgfsetstrokecolor{strokecol}
  \draw (329.5bp,670.5bp) node {0};
  \pgfsetcolor{red}
  \draw [->] (345.43bp,699.04bp) .. controls (363.62bp,685.04bp) and (390.99bp,663.96bp)  .. (419.45bp,642.05bp);
  \pgfsetstrokecolor{strokecol}
  \draw (393.5bp,670.5bp) node {1};
  \pgfsetcolor{blue}
  \draw [->] (318.92bp,609.39bp) .. controls (314.1bp,597.41bp) and (307.61bp,581.3bp)  .. (298.31bp,558.18bp);
  \pgfsetstrokecolor{strokecol}
  \draw (314.5bp,583.5bp) node {0};
  \pgfsetcolor{red}
  \draw [->] (333.89bp,609.39bp) .. controls (339.34bp,597.25bp) and (346.68bp,580.87bp)  .. (357.12bp,557.58bp);
  \pgfsetstrokecolor{strokecol}
  \draw (351.5bp,583.5bp) node {1};
  \pgfsetcolor{blue}
  \draw [->] (361.27bp,521.97bp) .. controls (358.83bp,510.19bp) and (355.6bp,494.56bp)  .. (350.72bp,471.0bp);
  \pgfsetstrokecolor{strokecol}
  \draw (361.5bp,496.5bp) node {0};
  \pgfsetcolor{red}
  \draw [->] (375.67bp,522.81bp) .. controls (383.28bp,510.54bp) and (393.67bp,493.8bp)  .. (407.8bp,471.05bp);
  \pgfsetstrokecolor{strokecol}
  \draw (399.5bp,496.5bp) node {1};
  \pgfsetcolor{blue}
  \draw [->] (439.0bp,608.97bp) .. controls (439.0bp,597.19bp) and (439.0bp,581.56bp)  .. (439.0bp,558.0bp);
  \pgfsetstrokecolor{strokecol}
  \draw (442.5bp,583.5bp) node {0};
  \pgfsetcolor{red}
  \draw [->] (452.88bp,610.23bp) .. controls (463.2bp,597.76bp) and (477.47bp,580.51bp)  .. (495.92bp,558.22bp);
  \pgfsetstrokecolor{strokecol}
  \draw (482.5bp,583.5bp) node {1};
  \pgfsetcolor{blue}
  \draw [->] (405.0bp,782.97bp) .. controls (405.0bp,771.19bp) and (405.0bp,755.56bp)  .. (405.0bp,732.0bp);
  \pgfsetstrokecolor{strokecol}
  \draw (408.5bp,757.5bp) node {0};
  \pgfsetcolor{red}
  \draw [->] (418.88bp,784.23bp) .. controls (429.2bp,771.76bp) and (443.47bp,754.51bp)  .. (461.92bp,732.22bp);
  \pgfsetstrokecolor{strokecol}
  \draw (449.5bp,757.5bp) node {1};
\begin{scope}
  \pgfsetstrokecolor{strokecol}
  \draw (102.0bp,210.0bp) -- (48.0bp,210.0bp) -- (48.0bp,174.0bp) -- (102.0bp,174.0bp) -- cycle;
  \draw (75.0bp,192.0bp) node {T};
\end{scope}
\begin{scope}
  \pgfsetstrokecolor{strokecol}
  \draw (192.0bp,36.0bp) -- (138.0bp,36.0bp) -- (138.0bp,0.0bp) -- (192.0bp,0.0bp) -- cycle;
  \draw (165.0bp,18.0bp) node {T};
\end{scope}
\begin{scope}
  \pgfsetstrokecolor{strokecol}
  \draw (264.0bp,36.0bp) -- (210.0bp,36.0bp) -- (210.0bp,0.0bp) -- (264.0bp,0.0bp) -- cycle;
  \draw (237.0bp,18.0bp) node {F};
\end{scope}
\begin{scope}
  \pgfsetstrokecolor{strokecol}
  \draw (149.0bp,123.0bp) -- (95.0bp,123.0bp) -- (95.0bp,87.0bp) -- (149.0bp,87.0bp) -- cycle;
  \draw (122.0bp,105.0bp) node {F};
\end{scope}
\begin{scope}
  \pgfsetstrokecolor{strokecol}
  \draw (54.0bp,297.0bp) -- (0.0bp,297.0bp) -- (0.0bp,261.0bp) -- (54.0bp,261.0bp) -- cycle;
  \draw (27.0bp,279.0bp) node {F};
\end{scope}
\begin{scope}
  \pgfsetstrokecolor{strokecol}
  \draw (202.0bp,297.0bp) -- (148.0bp,297.0bp) -- (148.0bp,261.0bp) -- (202.0bp,261.0bp) -- cycle;
  \draw (175.0bp,279.0bp) node {F};
\end{scope}
\begin{scope}
  \pgfsetstrokecolor{strokecol}
  \draw (274.0bp,297.0bp) -- (220.0bp,297.0bp) -- (220.0bp,261.0bp) -- (274.0bp,261.0bp) -- cycle;
  \draw (247.0bp,279.0bp) node {T};
\end{scope}
\begin{scope}
  \pgfsetstrokecolor{strokecol}
  \draw (276.0bp,384.0bp) -- (222.0bp,384.0bp) -- (222.0bp,348.0bp) -- (276.0bp,348.0bp) -- cycle;
  \draw (249.0bp,366.0bp) node {T};
\end{scope}
\begin{scope}
  \pgfsetstrokecolor{strokecol}
  \draw (348.0bp,297.0bp) -- (294.0bp,297.0bp) -- (294.0bp,261.0bp) -- (348.0bp,261.0bp) -- cycle;
  \draw (321.0bp,279.0bp) node {T};
\end{scope}
\begin{scope}
  \pgfsetstrokecolor{strokecol}
  \draw (420.0bp,297.0bp) -- (366.0bp,297.0bp) -- (366.0bp,261.0bp) -- (420.0bp,261.0bp) -- cycle;
  \draw (393.0bp,279.0bp) node {F};
\end{scope}
\begin{scope}
  \pgfsetstrokecolor{strokecol}
  \draw (174.0bp,558.0bp) -- (120.0bp,558.0bp) -- (120.0bp,522.0bp) -- (174.0bp,522.0bp) -- cycle;
  \draw (147.0bp,540.0bp) node {F};
\end{scope}
\begin{scope}
  \pgfsetstrokecolor{strokecol}
  \draw (172.0bp,645.0bp) -- (118.0bp,645.0bp) -- (118.0bp,609.0bp) -- (172.0bp,609.0bp) -- cycle;
  \draw (145.0bp,627.0bp) node {F};
\end{scope}
\begin{scope}
  \pgfsetstrokecolor{strokecol}
  \draw (318.0bp,558.0bp) -- (264.0bp,558.0bp) -- (264.0bp,522.0bp) -- (318.0bp,522.0bp) -- cycle;
  \draw (291.0bp,540.0bp) node {T};
\end{scope}
\begin{scope}
  \pgfsetstrokecolor{strokecol}
  \draw (374.0bp,471.0bp) -- (320.0bp,471.0bp) -- (320.0bp,435.0bp) -- (374.0bp,435.0bp) -- cycle;
  \draw (347.0bp,453.0bp) node {T};
\end{scope}
\begin{scope}
  \pgfsetstrokecolor{strokecol}
  \draw (446.0bp,471.0bp) -- (392.0bp,471.0bp) -- (392.0bp,435.0bp) -- (446.0bp,435.0bp) -- cycle;
  \draw (419.0bp,453.0bp) node {F};
\end{scope}
\begin{scope}
  \pgfsetstrokecolor{strokecol}
  \draw (466.0bp,558.0bp) -- (412.0bp,558.0bp) -- (412.0bp,522.0bp) -- (466.0bp,522.0bp) -- cycle;
  \draw (439.0bp,540.0bp) node {T};
\end{scope}
\begin{scope}
  \pgfsetstrokecolor{strokecol}
  \draw (538.0bp,558.0bp) -- (484.0bp,558.0bp) -- (484.0bp,522.0bp) -- (538.0bp,522.0bp) -- cycle;
  \draw (511.0bp,540.0bp) node {F};
\end{scope}
\begin{scope}
  \pgfsetstrokecolor{strokecol}
  \draw (432.0bp,732.0bp) -- (378.0bp,732.0bp) -- (378.0bp,696.0bp) -- (432.0bp,696.0bp) -- cycle;
  \draw (405.0bp,714.0bp) node {T};
\end{scope}
\begin{scope}
  \pgfsetstrokecolor{strokecol}
  \draw (504.0bp,732.0bp) -- (450.0bp,732.0bp) -- (450.0bp,696.0bp) -- (504.0bp,696.0bp) -- cycle;
  \draw (477.0bp,714.0bp) node {F};
\end{scope}
\begin{scope}
  \pgfsetstrokecolor{strokecol}
  \draw (364.0bp,888.0bp) ellipse (27.0bp and 18.0bp);
  \draw (364.0bp,888.0bp) node {1};
\end{scope}
\begin{scope}
  \pgfsetstrokecolor{strokecol}
  \draw (324.0bp,801.0bp) ellipse (29.5bp and 18.0bp);
  \draw (324.0bp,801.0bp) node {2};
\end{scope}
\begin{scope}
  \pgfsetstrokecolor{strokecol}
  \draw (405.0bp,801.0bp) ellipse (33.6bp and 18.0bp);
  \draw (405.0bp,801.0bp) node {35};
\end{scope}
\begin{scope}
  \pgfsetstrokecolor{strokecol}
  \draw (219.0bp,714.0bp) ellipse (27.0bp and 18.0bp);
  \draw (219.0bp,714.0bp) node {3};
\end{scope}
\begin{scope}
  \pgfsetstrokecolor{strokecol}
  \draw (326.0bp,714.0bp) ellipse (33.6bp and 18.0bp);
  \draw (326.0bp,714.0bp) node {26};
\end{scope}
\begin{scope}
  \pgfsetstrokecolor{strokecol}
  \draw (219.0bp,627.0bp) ellipse (29.5bp and 18.0bp);
  \draw (219.0bp,627.0bp) node {4};
\end{scope}
\begin{scope}
  \pgfsetstrokecolor{strokecol}
  \draw (219.0bp,540.0bp) ellipse (27.0bp and 18.0bp);
  \draw (219.0bp,540.0bp) node {5};
\end{scope}
\begin{scope}
  \pgfsetstrokecolor{strokecol}
  \draw (173.0bp,453.0bp) ellipse (29.5bp and 18.0bp);
  \draw (173.0bp,453.0bp) node {6};
\end{scope}
\begin{scope}
  \pgfsetstrokecolor{strokecol}
  \draw (249.0bp,453.0bp) ellipse (29.5bp and 18.0bp);
  \draw (249.0bp,453.0bp) node {19};
\end{scope}
\begin{scope}
  \pgfsetstrokecolor{strokecol}
  \draw (101.0bp,366.0bp) ellipse (27.0bp and 18.0bp);
  \draw (101.0bp,366.0bp) node {7};
\end{scope}
\begin{scope}
  \pgfsetstrokecolor{strokecol}
  \draw (175.0bp,366.0bp) ellipse (29.5bp and 18.0bp);
  \draw (175.0bp,366.0bp) node {16};
\end{scope}
\begin{scope}
  \pgfsetstrokecolor{strokecol}
  \draw (101.0bp,279.0bp) ellipse (29.5bp and 18.0bp);
  \draw (101.0bp,279.0bp) node {8};
\end{scope}
\begin{scope}
  \pgfsetstrokecolor{strokecol}
  \draw (149.0bp,192.0bp) ellipse (29.5bp and 18.0bp);
  \draw (149.0bp,192.0bp) node {10};
\end{scope}
\begin{scope}
  \pgfsetstrokecolor{strokecol}
  \draw (201.0bp,105.0bp) ellipse (33.6bp and 18.0bp);
  \draw (201.0bp,105.0bp) node {11};
\end{scope}
\begin{scope}
  \pgfsetstrokecolor{strokecol}
  \draw (323.0bp,366.0bp) ellipse (29.5bp and 18.0bp);
  \draw (323.0bp,366.0bp) node {21};
\end{scope}
\begin{scope}
  \pgfsetstrokecolor{strokecol}
  \draw (326.0bp,627.0bp) ellipse (33.6bp and 18.0bp);
  \draw (326.0bp,627.0bp) node {27};
\end{scope}
\begin{scope}
  \pgfsetstrokecolor{strokecol}
  \draw (439.0bp,627.0bp) ellipse (33.6bp and 18.0bp);
  \draw (439.0bp,627.0bp) node {32};
\end{scope}
\begin{scope}
  \pgfsetstrokecolor{strokecol}
  \draw (365.0bp,540.0bp) ellipse (29.5bp and 18.0bp);
  \draw (365.0bp,540.0bp) node {29};
\end{scope}
\end{tikzpicture}

}
    \caption{\texttt{postoperative-patient-data-un\_1-un} with 50\% sampling}\label{fig:patient}
\end{figure}
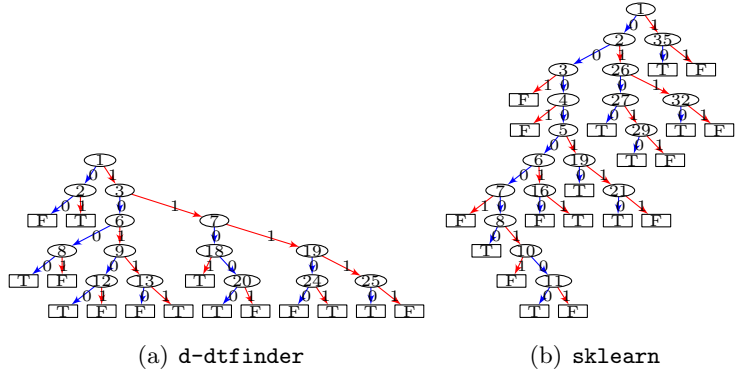

Overall, focusing on minimizing depth first is computationally advantageous, yet
yielding decision trees of good quality. We illustrate this on a particular instance.
 Figure~\ref{fig:patient} shows decision trees calculated by our approach  minimizing depth first (\ddtfinder) and  calculated by  the greedy approach (\sklearn).
  The optimal tree gives depth 6  and size 29, the greedy approach gives depth 11 and size 37.
  In contrast, the other approaches timeout on this instance in 1000~s.
\section{Related Work}\label{sec:Related}

Greedy algorithms for learning decision trees based on recursive splitting are
well-known~\cite{breiman-84,quinlan86,quinlan93}; see also~\cite{furnkranz2017}
for an overview.

Various notions of optimality of decision trees appear in the literature.  Some
approaches  focus on  finding a tree with a \emph{fixed depth} but with the
best \emph{accuracy}~\cite{verwer-aaai19,verhaeghe-bnaic19,bertsimas-ml17}.
These approaches assume a full (perfectly balanced) binary tree of the fixed
depth  whose accuracy is to be optimized.  While the problem is still very
hard, it is in some sense easier because the topology is fixed and only the
labeling needs to be calculated.  However, combinations of these approaches in
our approach is an interesting line of research.

Another approach is taken by~\cite{hu-nips19}, which optimizes a linear
combination of accuracy and size.  However, this approach is based on brute
force search  and in our experiments we were only able to synthesize trees with
a handful of features while the considered benchmarks contain hundreds of
features.

Closest to our work is~\cite{narodytska-ijcai18},  which uses SAT encoding to
construct a   size-optimal decision tree  for a given set of  consistent
samples.
In contrast to our work, individual
nodes and their children relation are modeled explicitly. This means that a
path from the root to a leaf is implicit.  In principle, one could also
restrict the depth of these implicit paths by adding additional counters or
some other form of cardinality constraints. This is bound to be less efficient.
Further, our encoding is closer to the idea of a tree.  If the
tree is modeled through nodes, it must be ensured that is in fact a tree via
cardinality constraints---ensuring that each node has  one and only one parent
(except for the root) and that  each  internal node has two children.  These
cardinality constraints are not needed in our encoding.  Since in our case,
classes are per path rather than node we save half of the semantic constraint
(see Section~\ref{sec:Encoding}).
It is interesting to  compare how symmetries are  broken
in~\cite{narodytska-ijcai18}, where restrictions are imposed on the possible children nodes.
In our approach
paths are ordered lexicographically %
rather than in an arbitrary order.  This order lets us single out the
leftmost in the rightmost branches, which turned out to be useful in
lower-bounding the depth (Section~\ref{sec:Optimizations}).  We   remark that
lexicographic order is a popular means of breaking symmetries in general
graphs, cf.~\cite{heule-mcs19}.

Earlier work for minimization of decision tree using Constraint Programming
(CP) exists~\cite{bessiere-cp09}.  It was shown in~\cite{narodytska-ijcai18},
that the approach by Narodytska~et~al.\ strictly outperforms the approach of
Bessiere~at~al.~this is most likely to be attributed to the fact  that the CP
encoding is asymptotically much larger.

Synthesis by calls to a SAT/SMT solver  has seen increased
interest in the recent years,
cf.~\cite{kolb-ijcai18,ignatiev-ijcar18,narodytska-sat19}.
Haaswi~et~al.\  used  topology enumeration to  synthesize  Boolean
circuits~\cite{haaswijk-dac18}.  The general idea is  analogous to our approach
(see  Section~\ref{sec:topologies}).  However, the set of possible  topologies
is partitioned differently.  The possible topologies are DAGs,  whereas they are
trees in our case.  Topologies in their approach belong to the same partition
if they have the same number of nodes at each level (levels are obtained by
BFS). This approach is unlikely to give good partitioning for binary trees
and is more expensive to encode  than our approach.  Further, in our approach,
the enumeration of topologies  simply goes over all possible topologies if the
number of nodes is small.

The well-known technique of \emph{cube-and-conquer} (CnC) splits the
search-space by a lookahead solver~\cite{hyvarinen-lpar10,heule-hvc11}.
The lookahead solver is run with a bound,
which yields cubes to be decided by a traditional CDCL solver.  Compared to our approach, CnC  is much
more general since it is applicable to any SAT instance, and, the lookahead
solver is less likely to generate cubes that will be decided trivially. The
downside is that CnC  may not come up with a splitting as a human would.
Further, the lookahead solver can be very costly.  In our preliminary
experiments, CnC performs much more poorly than a plain SAT solver on our
instances.  The order in which cubes are decided is also investigated by Heule~et~al.~\cite{heule-hvc11}.
 
\section{Conclusions  and  Future Work}\label{sec:Conclusions}

This paper proposes a novel  SAT-based encoding for decision trees,  which
enables natively controlling both the tree's size and depth.
We also study search-space splitting by topology enumeration.
Our implementation outperforms existing work of~\cite{narodytska-ijcai18}  but also enables a
finer control due to the explicit representation of paths of the tree.
This finer control lets us optimize practically interesting instances that had been out of reach.

The proposed approaches open a number of avenues for future research.  The
solving itself could be further improved by  better splitting, parallelization,
and combining with cube-and-conquer~\cite{heule-hvc11}.
While some  preprocessing of the examples was already
used in our optimization techniques (Section~\ref{sec:Optimizations}),
further inspection  could be used to draw more information from them, e.g.\
introduction of extended variables in the spirit of~\cite{biere-sat14}.
The proposed techniques  could also be integrated into more expressive approaches, e.g.\
SMT-based synthesis~\cite{kolb-ijcai18}.

At the application level, we are investigating the integration of  our tool with
some greedy approaches, e.g.\ ensembles, where only limited depth is
considered.  Or, consider a hybrid  between a greedy approach and an exact approach where an exact approach is
invoked on smaller sub-problems.   It would be interesting to
investigate  whether trees with a smaller  depth are really easier to
understand and interpret, and, what is the trade-off between depth and size.
Our approach provides the means to exactly quantify these metrics.

The experimental evaluation shows that SAT solvers poorly handle a search-space
with many topologies. We believe that this represents an important challenge
for the SAT community.

\paragraph{\bf Acknowledgements}
This work was supported by national funds through FCT, Funda{\c c}\~ao para a Ci\^encia e a Tecnologia, under project UIDB/50021/2020,
the project INFOCOS with reference PTDC/CCI-COM/32378/2017.
The results were  supported by the Ministry of Education,  Youth and Sports
within the dedicated program  ERC CZ  under  the project POSTMAN  with reference LL1902.

\end{document}